\documentclass[aps,prx,reprint,showpacs,eqsecnum]{revtex4-1}
\usepackage{graphicx}
\usepackage{amsmath}
\usepackage{amssymb}
\usepackage{amsfonts}
\usepackage{bbm}
\usepackage{bm}
\usepackage{verbatim}
\usepackage{cancel}
\usepackage[bookmarks=false,colorlinks=true,urlcolor=blue,citecolor=blue,linkcolor=blue]{hyperref}
\usepackage[normalem]{ulem}

\begin{document}

\title{Quantum Many-Body Scars and Space-Time Crystalline Order  \\ from Magnon Condensation}
%Quantum Many-Body Scars as a Magnon Tower with Space-Time Crystalline Order
%Magnon Tower, Space-Time Crystalline Order and Many-body Scars in a Rydberg Atom Chain}

\author{Thomas Iadecola}
\affiliation{Condensed Matter Theory Center and Joint Quantum Institute, Department of Physics, University of Maryland, College Park, Maryland 20742, USA}

\author{Michael Schecter}
\affiliation{Condensed Matter Theory Center and Joint Quantum Institute, Department of Physics, University of Maryland, College Park, Maryland 20742, USA}

\author{Shenglong Xu}
\affiliation{Condensed Matter Theory Center and Joint Quantum Institute, Department of Physics, University of Maryland, College Park, Maryland 20742, USA}

\date{\today}

\begin{abstract}
We study the eigenstate properties of a nonintegrable spin chain that was recently realized experimentally in a Rydberg-atom quantum simulator. In the experiment, long-lived coherent many-body oscillations were observed only when the system was initialized in a particular product state. This pronounced coherence has been attributed to the presence of special ``scarred'' eigenstates with nearly equally-spaced energies and putative nonergodic properties despite their finite energy density.  In this paper we uncover a surprising connection between these scarred eigenstates and low-lying quasiparticle excitations of the spin chain.  In particular, we show that these eigenstates can be accurately captured by a set of variational states containing a macroscopic number of magnons with momentum $\pi$.  This leads to an interpretation of the scarred eigenstates as finite-energy-density condensates of weakly interacting $\pi$-magnons.  One natural consequence of this interpretation is that the scarred eigenstates possess long-range order in both space and time, providing a rare example of the spontaneous breaking of \emph{continuous} time-translation symmetry.  We verify numerically the presence of this space-time crystalline order and explain how it is consistent with established no-go theorems precluding its existence in ground states and at thermal equilibrium.
\end{abstract}

\maketitle

\tableofcontents

\section{Introduction}

A recent experiment~\cite{Bernien17} in Rydberg-blockaded atomic lattices showed unexpected persistent oscillations in measurements of local observables after a quench from the N\'eel state.  
The existence of such oscillations and their long-lived coherence is remarkable given that the chosen initial state nominally resembles an infinite-temperature state from the point of view of the underlying quantum many-body system, which is strongly interacting.  In such a scenario, the expectation based on the eigenstate thermalization hypothesis (ETH)~\cite{Deutsch91,Srednicki94,D'Alessio16,Deutsch18} is that the system rapidly loses memory of the initial state and becomes essentially featureless.  Although it is well known that such ergodic dynamics can be avoided due to either integrability or many-body localization (MBL)~\cite{Polkovnikov11,Nandkishore15,Altman15,Abanin18}, neither of these scenarios appears to apply for the experiment in question. These experimental results therefore present an interesting and fundamental puzzle that has attracted substantial recent attention.

The observed oscillations were quickly shown to result from the existence of a set of special ``scarred'' eigenstates~\cite{Turner17,Turner18,Choi18} that have anomalously high overlap with the N\'eel state, anomalously low entanglement, and a near-constant energy spacing between them, giving rise to long-lived quantum coherence.  While various properties of these ``quantum many-body scar" states have been established~\cite{Turner17,Turner18,Choi18,Khemani18,Lin18,Ho19}, a microscopic picture of their origin is lacking. This leaves open a number of important questions regarding their robustness, ubiquity and physical interpretation.

The surge of interest in these eigenstates has also drawn attention to earlier examples of ``special" low-entanglement eigenstates in otherwise generic models, including the ``$\eta$-pairing'' states in the Hubbard model~\cite{Yang89,Vafek17} and the bimagnon tower of states in nonintegrable AKLT spin chains~\cite{Moudgalya18a,Moudgalya18b}~\footnote{It is worth pointing out that the $\eta$-pairing states do not violate the strong ETH as we allude to it in Sec.~\ref{sec:model}, while the AKLT bimagnon states do.  The reason is that the $\eta$-pairing states happen to be the only states in their symmetry sector, and thus their entanglement can be at most logarithmic.}. These examples are appealing due to their analytical exactness and clear underlying physical picture, but it has hitherto not been clear to what extent, if at all, such a picture underlies the scarred eigenstates of the model simulated by the experiment~\cite{Bernien17}. 

In this paper we demonstrate that the scarred eigenstates found in Ref.~\cite{Turner17} admit a surprising description in terms of low-lying magnon excitations above the paramagnetic ground state of the experimentally relevant model.  In particular, we show that a basis of only polynomially many variational states built primarily of magnons carrying momentum $\pi$ captures both zero- and finite-energy-density scar states with remarkable accuracy.  We further demonstrate that the scar states at finite energy density harbor a macroscopic number of these $\pi$-magnons, suggesting an interpretation of the scar states as $\pi$-magnon condensates.  This allows us to make contact with the aforementioned $\eta$-pairing states in the Hubbard model and bimagnon states in the AKLT models, which also feature macroscopic numbers of $\pi$-momentum excitations.

The $\pi$-magnon-condensate picture proposed here implies that the scar states exhibit long-range order at wavenumber $\pi$.  We show that this in combination with the near-constant spacing of energy levels leads to unequal-time connected correlations that are periodic in both space and time in the scarred eigenstates at finite energy density.  Such nontrivial unequal-space-time connected correlators are a definitive~\cite{Watanabe15} hallmark of space-time crystalline order~(see Ref.~\cite{Sacha17} for a review). Notably, the space-time crystalline order reported here involves the spontaneous breaking of \emph{continuous} time-translation symmetry and thus more closely resembles the original proposal for time crystals~\cite{Wilczek12} than it does the so-called ``discrete time crystal" scenario~\cite{Sacha15,Khemani16,Else16,vonKeyserlingk16} that has been observed in several recent experiments on periodically driven systems~\cite{Choi17,Zhang17,Pal18,Rovny18a,Rovny18b,Smits18}.  Such order is enabled without periodic driving in the present context by the highly-excited yet nonthermal nature of scar states at finite energy density. This spatiotemporal long-range order is an unusual example of eigenstate order~\cite{Parameswaran17}, which is typically only associated with MBL systems but here arises in the unorthodox setting of rare eigenstates in an otherwise ergodic quantum system.

The remainder of this paper is organized as follows. In Sec.~\ref{sec:model} we introduce the Rydberg-atom model studied experimentally in Ref.~\cite{Bernien17} and briefly review the phenomena associated with quantum many-body scars. In Sec.~\ref{sec:sma} we study the low-energy properties of the Rydberg model using the single-mode approximation (SMA) and motivate the possibility of a $\pi$-magnon description of the scar
states. In Sec.~\ref{sec:tower} we validate this description by showing how the SMA can be extended to build a tower of $\pi$-magnon states whose span captures the scarred eigenstates with high fidelity. We explore the consequences of the $\pi$-magnon description in Sec.~\ref{sec:order} and show that it implies long-range space-time crystalline order of the scar states at finite energy density. 
In Sec.~\ref{sec:deform} we show that a previously studied deformation of the model that was found to enhance the stability of the observed oscillatory dynamics~\cite{Khemani18,Choi18} also enhances the $\pi$-magnon description and long-range order of the scar states, further emphasizing the relevance of our picture.
Concluding remarks and discussion are presented in Sec.~\ref{sec:conclusion}.

\section{Model}\label{sec:model}

%%%%%%%%%%%%
%%%%%%%%%%%%
\begin{figure}[t!]
\begin{center}
\includegraphics[width=\columnwidth]{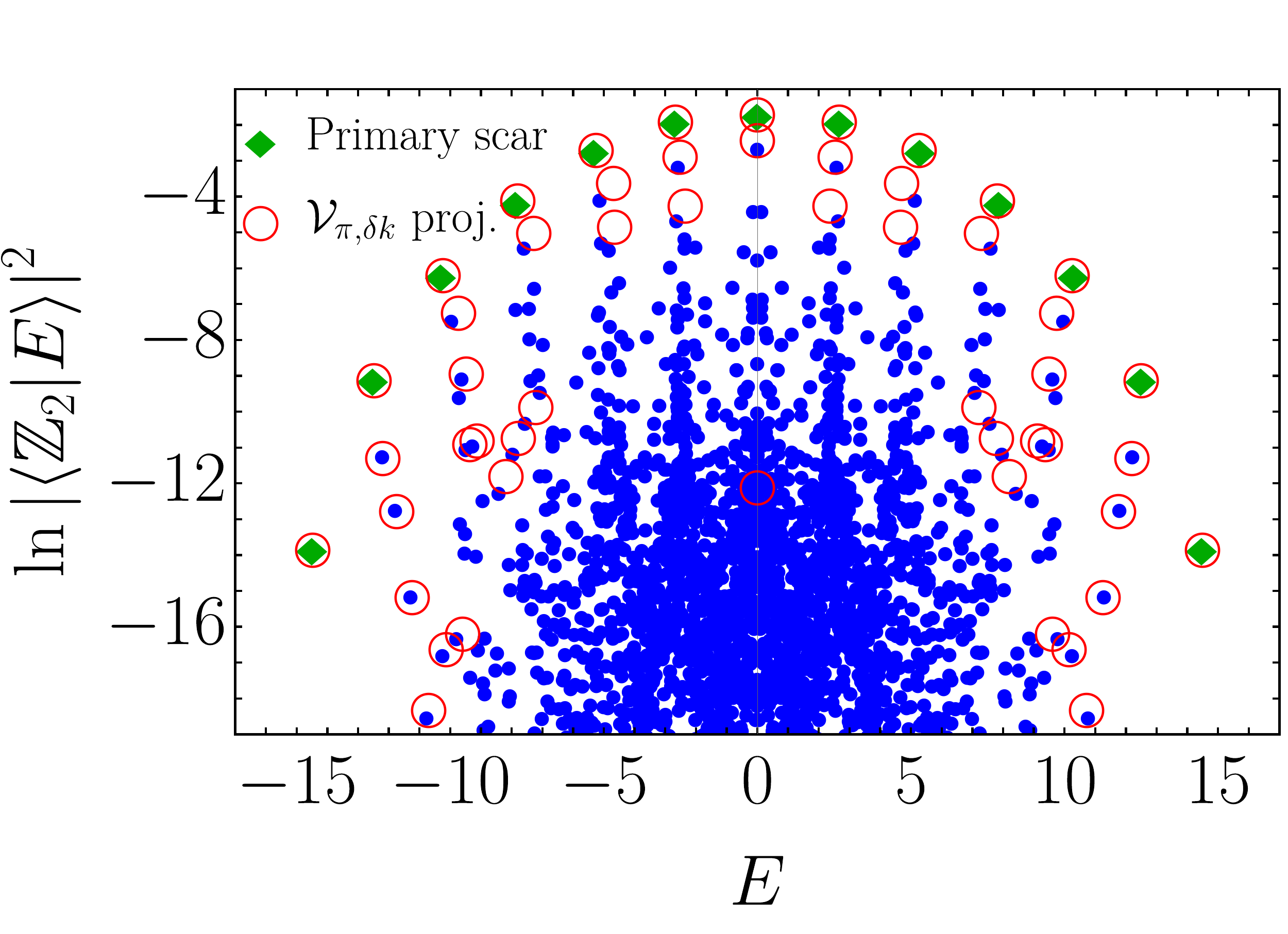}
\caption{Overlaps with the N\'eel state $|\mathbb Z_2\rangle$ of the eigenstates of \eqref{eq:H} as a function of their energy $E$ (blue dots). The data shown are for $L=24$ in the $k=0$ sector. The primary scar states are denoted by green diamonds. The circles denote the $\mathbb Z_2$ overlap and energy expectation values of the (renormalized) \emph{projections} of eigenstates onto the multimagnon vector space $\mathcal V_{\pi,\delta k}$ (discussed in Sec.~\ref{sec:tower}). Only the projections of eigenstates whose total weight in $\mathcal V_{\pi,\delta k}$ exceeds 60\% are shown. The multimagnon vector space captures well not only the primary scar states, but multiple ``descendant" scar states at the same energy density that have smaller N\'{e}el overlap.}
\label{fig:overlap}
\end{center}
\end{figure}
%%%%%%%%%%%%
%%%%%%%%%%%%

In this paper we focus on an effective model that is relevant to ongoing experiments studying arrays of Rydberg atoms \cite{Bernien17,Lienhard18,Guardado-Sanchez18}. As discussed in Refs.~\cite{Lesanovsky11,Bernien17,Turner17}, the nearest-neighbor Rydberg blockade regime in such arrays is well-described by the so-called ``PXP'' model, which can be written in spin-1/2 notation as
\begin{equation}
\label{eq:H}
H=\sum_{j=1}^L P_{j-1}X_j P_{j+1},
\end{equation}
where $L$ is the system length,
\begin{align}
P_j=\frac{\mathbbm{1}-Z_j}{2}
\end{align}
is a local projector onto spin down, and where we denote the Pauli operators on site $j$ by $X_j$, $Y_j$, and $Z_j$.  Throughout this work we assume periodic boundary conditions such that $j=L+1\equiv 1$. The ``up" and ``down" spin states $|\!\!\uparrow\,\rangle$ and $|\!\!\downarrow\,\rangle$ represent a Rydberg atom in its excited and ground state, respectively. The Hamiltonian \eqref{eq:H} is thus interpreted as a sum of local spin-flip operators resulting from driving atoms between their ground and excited states, subject to the additional Rydberg-blockade constraint that no two neighboring atoms can be excited simultaneously due to strong nearest-neighbor interactions. The dimension of the Hilbert space in the presence of the blockade constraint scales as $\varphi^L$, where $\varphi=1.618\dots$ is the golden ratio.

We now review several remarkable properties of the model \eqref{eq:H} that have previously been identified. It was first shown in the experiment~\cite{Bernien17} that a quench from the N\'{e}el state,
\begin{align}
|\mathbb Z_2\rangle=|\!\uparrow\downarrow\uparrow\downarrow\cdots\rangle,
\end{align}
leads to long-lived persistent oscillations of local observables far beyond the $O(1)$ microscopic timescale of Eq.~\eqref{eq:H}. It was later shown that this phenomenon may be attributed to an extensive (but measure-zero) set of special ``scarred'' eigenstates that have nearly equal energy spacing, unusually high overlap with the N\'{e}el state (see, e.g.,~Fig.~\ref{fig:overlap}) and anomalously low entanglement scaling as $\ln L$~\cite{Turner17,Turner18,Choi18}.  We call the scar states having the highest overlap with the N\'eel state at a given energy density the ``primary" scar states. There are exactly $L+1$ such states, and their properties were found in Refs.~\cite{Turner17,Turner18} to be well captured by a forward scattering approximation (FSA) inspired by Krylov subspace techniques. The symmetry structure of the primary scar states, which will be important for the discussion below, is such that their total momentum (either 0 or $\pi$) alternates successively as their energy increases from the ground state (which is a zero-momentum scar state).

The existence of such special eigenstates in the model~\eqref{eq:H} directly contradicts~\cite{Lin18} the strong ETH, which states that \emph{every} eigenstate at a given finite energy density yields expectation values of local observables that are consistent with the canonical ensemble at a temperature determined by the energy density~\cite{Deutsch91,Srednicki94,D'Alessio16,Deutsch18}. For states in the middle of the many-body spectrum (nominally at infinite temperature), the ETH predicts volume-law entangled states, whose entanglement scales as $L$, in contrast to scarred states, whose entanglement evidently scales as $\ln L$~\cite{Turner17,Turner18,Choi18}. The strong ETH also strictly forbids long-range order of any type at infinite temperature, since the thermal density matrix, $\hat\rho=e^{-\beta H}$, is trivial at $\beta=0$. We stress that the existence of scarred eigenstates (which indeed have long-range order at finite energy density, as we show below) is perfectly consistent with the notion of nonintegrability and the \emph{weak} ETH, which allows a zero-measure set of nonthermal states~\cite{Deutsch18}.

The effect on the scar states of deformations of the Hamiltonian \eqref{eq:H} was studied in Refs.~\cite{Khemani18,Choi18}. Ref.~\cite{Khemani18} found that a particular deformation enhances the oscillatory dynamics of local observables.  In Ref.~\cite{Choi18} it was shown that a generalized version of this deformation can be used to systematically improve the accuracy of the FSA. These observations seem to suggest a notion of robustness for the scarring phenomenon. Enhancement of the many-body scars under the Hamiltonian deformation has been attributed to the emergence of an approximate SU$(2)$ algebra~\cite{Choi18} that, if exact, would enforce revivals in the dynamical evolution of the N\'{e}el state.

Motivated by the archetypal analytically exact scarred eigenstates in the Hubbard and AKLT models, in this paper we take an alternative ``bottom-up'' approach to the question of many-body quantum scars in the PXP model. In particular, we study whether and how the low-energy excitations of the model \eqref{eq:H} can be used to construct scarred eigenstates at finite energy density. We find that such a construction is indeed possible and that the scarred eigenstates of Eq.~\eqref{eq:H} can be well approximated by states in which a variable number of magnons with momentum $\pi$ are created above the ground state, see Fig.~\ref{fig:overlap}.

\section{Motivating the $\pi$-Magnon Description: Low-Lying States of the PXP Model}\label{sec:sma}

In this section we analyze some properties of the low-energy states of the Hamiltonian \eqref{eq:H} and motivate the $\pi$-magnon description of the scar states.

\subsection{Ground State}

The ground state of the PXP model is gapped and features exponentially decaying correlations of local observables~\cite{Ovchinnikov03}. Although it is not known analytically (unlike in related Rydberg-relevant models~\cite{Lesanovsky12a}), the ground state may nevertheless be studied for large systems using DMRG or expressed rather accurately via a variational wavefunction~\cite{Ovchinnikov03}
\begin{equation}\label{eq:ground}
|\psi_0\rangle=\sum_{m=0}^{L/2}\frac{c_m}{\sqrt{w_m}}(P\sigma^+)^m|\downarrow\downarrow\cdots\rangle,
\end{equation}
where $\sigma^+=\frac{1}{2}\sum_j (X_j+\mathrm i\, Y_j)$ is a spin raising operator, $P$ is a global projector onto the constrained Hilbert space, $w_m=\frac{(L-m-1)!L}{m!(L-2m)!}$ is a degeneracy factor and $c_m\propto e^{-\frac{(m-m_0)^2}{2\delta^2}}$ is approximately Gaussian. The wavefunction \eqref{eq:ground} was shown in Ref.~\cite{Ovchinnikov03} to yield estimates for the ground-state energy and magnetization densities in good quantitative agreement with DMRG. The squared overlap of this state with the true ground state of Eq.~\eqref{eq:H} at $L=24$ is roughly $0.98$. 

\subsection{Single-Mode Approximation}\label{sec:sma_b}

Excitations on top of the ground state can be generated using the single-mode approximation (SMA)~\cite{Bijl40,Feynman54,Girvin86}, where one acts on the ground state with a density operator carrying momentum $k$. We utilize the $z$-component of spin to express SMA states as
\begin{subequations}\label{eq:zk}
\begin{equation}
|k\rangle=\mathcal N_k\, Z_k|{\rm GS}\rangle,
\end{equation}
where $|{\rm GS}\rangle$ is the true ground state of Eq.~\eqref{eq:H},
\begin{align}\label{eq:Z_k}
Z_k = \sum_j e^{\mathrm ijk}Z_j,
\end{align}
and 
\begin{align}
\mathcal N_k=\langle Z_{-k}Z_k\rangle_0^{-1/2},
\end{align}
\end{subequations}
where 0 denotes a ground state expectation value. The excitation spectrum, $E_k=\langle k|H|k\rangle$, may be expressed as
\begin{align}\label{eq:Ek}
E_k-E_0=\frac{\langle [Z_{-k},[H,Z_k]]\rangle_0}{2\langle Z_{-k}Z_k\rangle_0}=\frac{-2E_0}{\langle Z_{-k}Z_k\rangle_0},
\end{align}
where $E_0=-0.6034\, L$ is the ground state energy~\cite{Ovchinnikov03} and where we have explicitly evaluated the double commutator using Eq.~\eqref{eq:H} and \eqref{eq:Z_k}.
Eq.~\eqref{eq:Ek} resembles the SMA excitation spectrum for the famous examples of superfluid Helium and the fractional quantum Hall effect~\cite{Bijl40,Feynman54,Girvin86}.

\begin{figure}[t!]
\begin{center}
\includegraphics[width=\columnwidth]{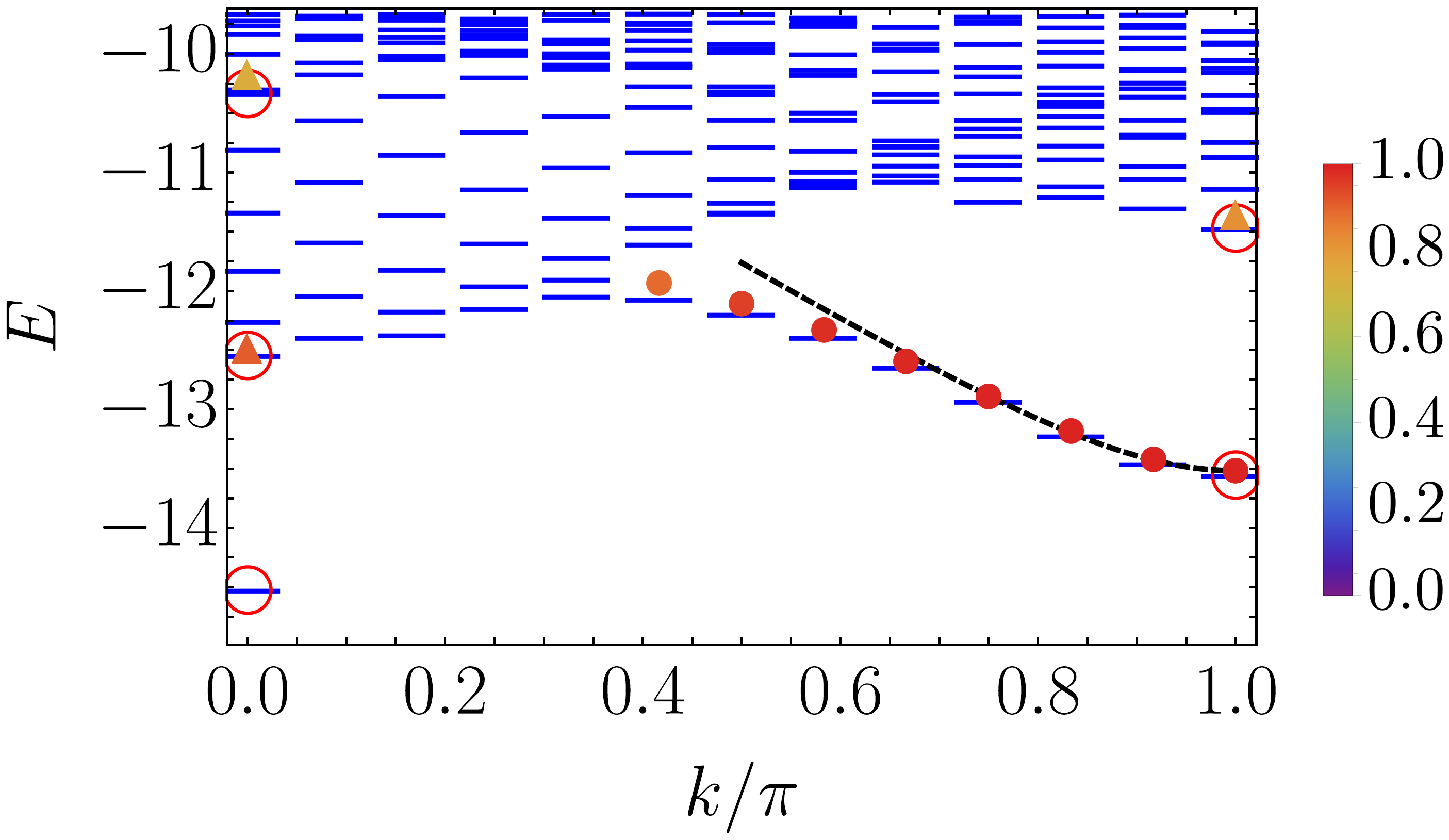}
\caption{Low-lying energy levels (blue ticks) of Eq.~\eqref{eq:H} as a function of total momentum $k$ for $L=24$. The SMA energies given by Eq.~\eqref{eq:Ek} are plotted as circles whose color scale indicates the overlap of the corresponding SMA states with exact eigenstates. A single-magnon branch with a minimum at $k=\pi$ is clearly visible, and eventually merges with the two-magnon continuum near $k=0.4\pi$. The dashed line is a fit to $\sqrt{\Delta^2+v^2k^2}$ near $k=\pi$. Primary scar states, which exist only at momenta $k=0,\pi$, are highlighted by red circles. The higher primary scar states can be well represented by a basis of $\pi$-magnon states (triangles) as we discuss in Sec.~\ref{sec:tower}.}
\label{fig:sma}
\end{center}
\end{figure}

The first excited state, which we denote $|1\rangle$, has momentum $k=\pi$ and is gapped by an amount $\Delta=0.9682$ \cite{Ovchinnikov03}, which is within 0.45\% of the extrapolated value of Eq.~\eqref{eq:Ek}. We plot the low-lying spectrum of Eq.~\eqref{eq:H} in Fig.~\ref{fig:sma}, where a single-magnon branch with a minimum at $k=\pi$ is clearly visible.  Fitting the single-magnon branch near $k=\pi$ to a relativistic spectrum $E_k-E_0=\sqrt{\Delta^2+v^2 k^2}$ and performing finite-size scaling, we find $v=1.6985$, implying a ground-state correlation length of $\xi=v/\Delta=1.7542$. One can see from Fig.~\ref{fig:sma} that the SMA branch eventually terminates around $k=0.4\pi$ where it merges with the two-magnon continuum.

%%%%%%%%%%%%
%%%%%%%%%%%%
%%%%%%%%%%%%
%%%%%%%%%%%%

We highlight in Fig.~\ref{fig:sma} the location of the low-lying primary scar states, which have anomalously high overlap with the $|\mathbb Z_2\rangle$ state. One sees that the scar states live only at momenta $k=0,\pi$ and include, in particular, the ground and first excited states. Since the ground and first excited states can be viewed as states with zero and one $\pi$-magnon, respectively, we speculate that higher-energy scar states may be obtained by repeated creation of $\pi$-magnons above the ground state. This is motivated by the fact that such a construction (a) would naturally lead to the observed alternation of the total momentum between $k=0$ and $k=\pi$ from one scar state to the next, and (b) would give rise to states with entanglement entropy scaling as $\ln L$~\cite{Turner17,Turner18} and progressively higher overlap with the $|\mathbb Z_2\rangle$ state, due to the repeated application of the local SMA operator $Z_\pi$ on the area-law-entangled ground state.

\subsection{Variational $\pi$-Magnon Creation Operator}\label{sec:creation}

\begin{figure}[t!]
\begin{center}
\includegraphics[width=0.8\columnwidth]{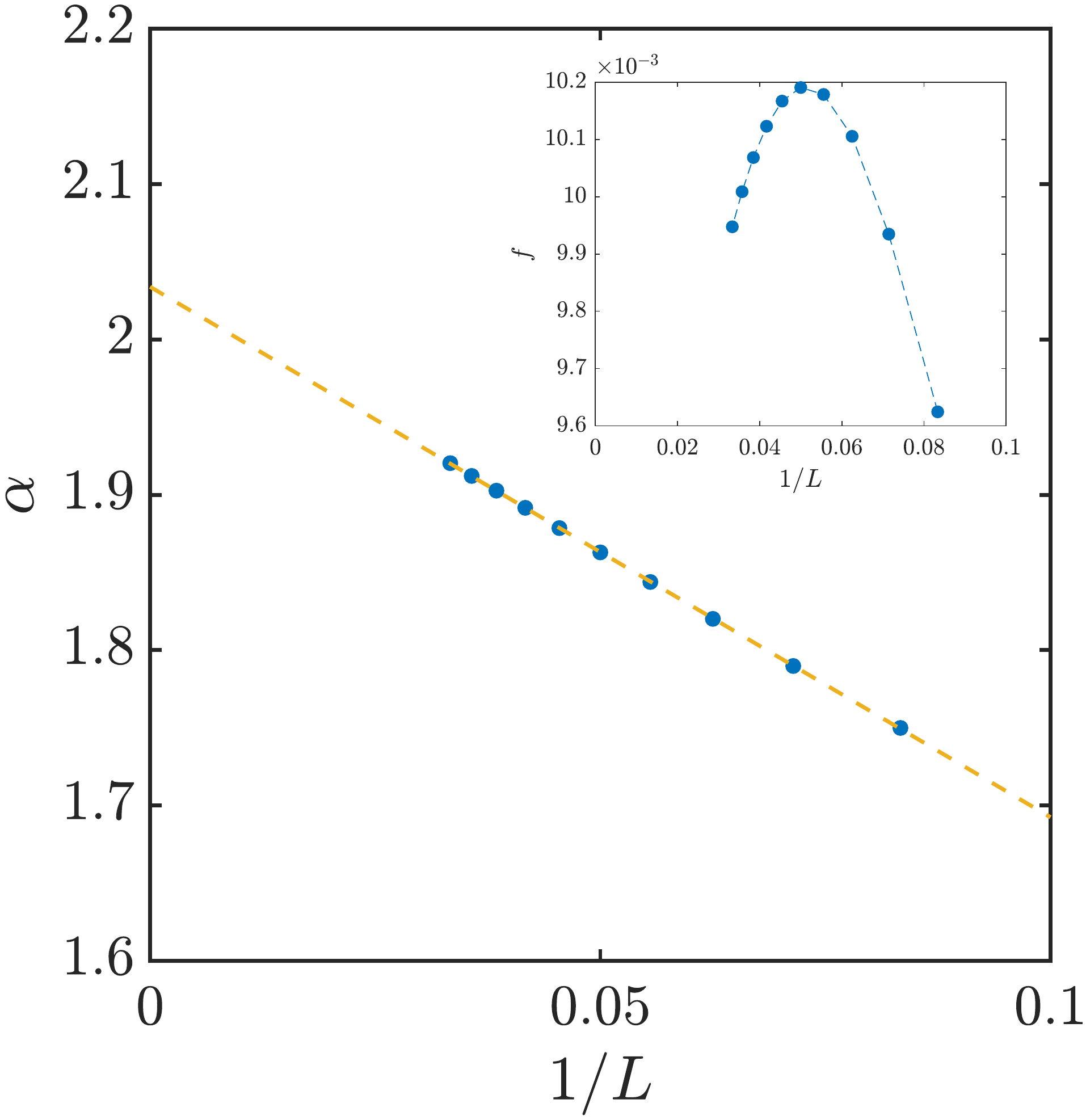}
\caption{Optimal coefficient $\alpha$ of Eq.~\eqref{eq:s+-} as a function of $1/L$ for $L=12,\dots,30$. We determine $\alpha$ by minimizing the cost function~\eqref{eq:f}. Linear extrapolation to $L=\infty$ yields $\alpha=2.03$. Inset: Optimal value of $f$ as a function of $1/L$.  For large $L$ the optimal value of $f$ decreases monotonically.}
\label{fig:a}
\end{center}
\end{figure}

These considerations motivate us to determine the $\pi$-magnon creation and annihilation operators $S^{\pm}_\pi$ to be used in constructing states with arbitrary numbers of $\pi$-magnons. We do this by combining SMAs for both the $z$ and $y$ spin projections,
\begin{subequations}
\begin{equation}\label{eq:s+-}
S^{\pm}_\pi=\frac{Z_{\pi}\mp \mathrm i\, \alpha Y_{\pi}}{2},
\end{equation}  
where $Y_\pi$ is understood to act only in the projected subspace, i.e.,
\begin{align}
Y_\pi=\sum_j (-1)^j\, P_{j-1}Y_j P_{j+1}.
\end{align}
\end{subequations}
The real coefficient $\alpha$ is a variational parameter that we determine by numerical minimization of the positive-semidefinite cost function
\begin{align}\label{eq:f}
f(\alpha)\!=\!1\!-\!\frac{1}{2}\left(\frac{|\langle 1|S^+_\pi |{\rm GS}\rangle|^2}{|\langle {\rm GS}|S^-_\pi S^+_\pi |{\rm GS}\rangle|}\!+\!\frac{|\langle {\rm GS}|S^-_\pi |1\rangle|^2}{|\langle 1|S^+_\pi S^-_\pi |1\rangle|}\right),
\end{align}�
where $|1\rangle$ is the first excited state. The cost function $f(\alpha)=0$ when both $S^+_\pi|{\rm GS}\rangle\propto|1\rangle$ and $S^-_\pi|1\rangle\propto|{\rm GS}\rangle$, i.e.~when $S^{\pm}_\pi$ act as perfect creation and annihilation operators for a $\pi$-magnon.
We plot the optimal value of $\alpha$ as a function of $1/L$ in Fig.~\ref{fig:a}, where it is shown that linear extrapolation to $L=\infty$ gives
$\alpha = 2.03$. In Sec.~\ref{sec:su(2)} we show that $\alpha\approx 2$ has an interesting interpretation in terms of an approximate SU$(2)$ algebra generated by $S^\pm_\pi$ and $H$.

%%%%%%%%%%%%
%%%%%%%%%%%%
%%%%%%%%%%%%
%%%%%%%%%%%%

The form of Eq.~\eqref{eq:s+-} can be fixed by considering, for simplicity, only single-body operators that act in the projected space. This restricts us to $Z_\pi,Y_\pi,X_\pi$ whose coefficients must be chosen to respect time-reversal symmetry (which acts as complex conjugation). At the same time, the Hamiltonian~\eqref{eq:H} has a spectral reflection symmetry,
\begin{subequations}
\begin{align}\label{eq:reflection}
CHC=-H,
\end{align} generated by
\begin{align}\label{eq:C}
C=\prod_i Z_i,
\end{align}
\end{subequations}
that exchanges eigenstates with positive and negative energies.  In particular, it maps the ground state $|{\rm GS}\rangle$ to the maximal-energy or ``ceiling state" $|{\rm CS}\rangle\equiv C|{\rm GS}\rangle$. Spectral reflection symmetry thus forces the lowering operator $S^-_\pi$ to be related to the raising operator as
\begin{align}
C S^{\pm}_\pi C=S^{\mp}_\pi.
\end{align}
Together with time-reversal symmetry, this implies that the coefficient of $X_\pi$ must vanish, hence Eq.~\eqref{eq:s+-}.
 
Using the extrapolated optimized value $\alpha=2.03$ at $L=24$ we obtain $\frac{|\langle 1|S^+_\pi |{\rm GS}\rangle|^2}{|\langle {\rm GS}|S^-_\pi S^+_\pi |{\rm GS}\rangle|}=0.988\dots$ and $\frac{|\langle {\rm GS}|S^-_\pi |1\rangle|^2}{|\langle 1|S^+_\pi S^-_\pi |1\rangle|}=0.989\dots$, indicating that the operators $S^{\pm}_\pi$ with $\alpha = 2.03$ indeed act as approximate $\pi$-magnon creation and annihilation operators.  Hereafter we set $\alpha=2$ in Eq.~\eqref{eq:s+-} for simplicity. Systematic improvements to Eq.~\eqref{eq:s+-} could be obtained by including longer-range multibody operators and more variational parameters, or by employing matrix product state techniques~\cite{Vanderstraeten14}.

 \subsection{Nearly-SU$(2)$ Algebra}\label{sec:su(2)}
 
 The value $\alpha\approx 2$ obtained numerically in Sec.~\ref{sec:creation} has an interesting interpretation in light of the algebra generated by the ladder operators~\eqref{eq:s+-} and the Hamiltonian~\eqref{eq:H}.  This algebra is given by
\begin{subequations}\label{eq:com1}
 \begin{align}
 \begin{split}
 [S^+_\pi,S^-_\pi]&=\alpha H\\
 [H,S^\pm_\pi]&=\pm\left(\frac{\alpha}{4}Z_\pi\mp \mathrm i\, Y_\pi\right)\pm O_{ZZZ},
 \end{split}
 \end{align}
where
\begin{align}
O_{ZZZ}=\frac{\alpha}{4}\sum_j (-1)^j\, Z_{j-1}Z_j Z_{j+1}.
\end{align}
\end{subequations}
If the term $O_{ZZZ}$ were not present, setting $\alpha=2$ would yield $[H,S^\pm_\pi]= S^\pm_\pi$, so that $S^\pm_\pi$ would be exact ladder operators with respect to the Hamiltonian.  If this were the case, and if we had $S^-_\pi|{\rm GS}\rangle=0$, then one would obtain an exact tower of $L+1$ $\pi$-magnon eigenstates proportional to $(S^+_\pi)^n|{\rm GS}\rangle$ for $n=0,\dots,L$.  Moreover, the $\pi$-magnon number operator $S^+_\pi S^-_\pi$ would become a conserved quantity.

The presence of the term $O_{ZZZ}$ spoils the otherwise closed algebra and appears to preclude the existence of an exact SU$(2)$ algebra associated with ladder operators $S^\pm_\pi$ of the form \eqref{eq:s+-}.  Nevertheless, the fact that the numerically determined optimal value $\alpha\approx 2$ in Eq.~\eqref{eq:s+-} indicates that, at least at low energies, the presence of the operator $O_{ZZZ}$ does not drastically alter the excitation spectrum in the thermodynamic limit.

\subsection{Relation to FSA}

A nearly-SU$(2)$ algebra equivalent to Eqs.~\eqref{eq:com1} appears in Ref.~\cite{Choi18} in the context of the FSA description of the scarred eigenstates.  However, the algebra in that work treats the Hamiltonian \eqref{eq:H} itself as a sum of raising and lowering operators for the staggered magnetization operator $Z_\pi$.  Thus, it is tempting to view the algebra of Eqs.~\eqref{eq:com1} as being related to the algebra of Ref.~\cite{Choi18} by a change of basis.  While this is conceptually accurate, it is not quantitatively accurate at the level of the $\pi$-magnon basis constructed in Sec.~\ref{sec:tower_a} and the FSA basis constructed in Refs.~\cite{Turner17,Turner18,Choi18}.  The source of quantitative discrepancy between the two approaches is precisely the presence of the operator $O_{ZZZ}$ that precludes an exact SU$(2)$ rotation between the two bases.  It is natural to speculate that the two approaches would be related by such a transformation should there exist a point in parameter space where the algebra \eqref{eq:com1} becomes an exact SU$(2)$ algebra, as suggested in Ref.~\cite{Choi18}.

We nevertheless emphasize the important difference between the two approaches in terms of physical intuition: whereas the FSA basis states consist of variable numbers of spin flips on top of the N\'eel state, the $\pi$-magnon states constructed in Sec.~\ref{sec:tower_a} consist of macroscopic numbers of $\pi$-magnons on top of the ground state of Eq.~\eqref{eq:H}.  The latter states therefore have a clear physical interpretation in terms of low-lying excitations of the model, whereas the former states do not have an obvious \emph{a priori} relationship to eigenstates of $H$ and rather span a Krylov subspace used to diagonalize it.  The $\pi$-magnon picture proposed in this work also motivates systematic improvements to the $\pi$-magnon basis constructed in Sec.~\ref{sec:tower_a}, which we discuss in Sec.~\ref{sec:tower_b}.
  
\section{Testing the $\pi$-Magnon Description}\label{sec:tower}

\subsection{$\pi$-Magnon Tower Basis}\label{sec:tower_a}

Despite the fact that the variational operators $S^\pm_\pi$ do not act as exact SU$(2)$ ladder operators with respect to the Hamiltonian~\eqref{eq:H}, we nevertheless can test the extent to which the tower of $\pi$-magnon states proposed in Sec.~\ref{sec:sma_b} is capable of describing the scarred eigenstates.  To this end, we define the ``$n$-$\pi$-magnon" tower of states
\begin{align}
\label{eq:nstates}
|n\rangle = \mathcal N_n\, (S^+_\pi)^n|{\rm GS}\rangle,
\end{align}
where $\mathcal N_n$ is a normalization factor and $S^+_\pi$ is given by Eq.~\eqref{eq:s+-} with $\alpha=2$.  In order to capture the spectral-reflection symmetry of $H$, and by extension that of the set of scar states, we will also find it useful to consider the ``reflected" $\pi$-magnon tower of states
\begin{align}
\label{eq:ntilstates}
|\tilde n\rangle = \mathcal N_{n}\, (S^-_\pi)^n|{\rm CS}\rangle=C|n\rangle,
\end{align}
where the ceiling state $|{\rm CS}\rangle=C|{\rm GS}\rangle$ and where the operator $C$ is defined in Eq.~\eqref{eq:C}.  Note that the towers of states \eqref{eq:nstates} and \eqref{eq:ntilstates} automatically include the ground and ceiling states, as $|0\rangle\equiv|{\rm GS}\rangle$ and $|\tilde 0\rangle\equiv|{\rm CS}\rangle$.

The states \eqref{eq:nstates} and \eqref{eq:ntilstates} have a straightforward physical interpretation.  The former is a state in which some number $n$ of magnons carrying momentum $\pi$ are created atop the ground state; the latter is an analogous state with respect to the ceiling state. However, these states by themselves cannot be true eigenstates of $H$.  In particular, the set $\{|n\rangle\}$ is not orthogonal, nor is the set $\{|\tilde n\rangle\}$; moreover, the states in $\{|n\rangle\}$ are not orthogonal to those in $\{|\tilde n\rangle\}$.  This failure of orthogonality is a consequence of the fact that $S^\pm_\pi$, which are essentially determined variationally, are not perfect ladder operators.  Nevertheless, it is still meaningful to ask whether the exact scarred eigenstates can be accurately represented in terms of the $n$-$\pi$-magnon states \eqref{eq:nstates} and their reflected counterparts \eqref{eq:ntilstates}.   

To investigate this possibility, we implement the following protocol.  Given an eigenstate $|E\rangle$ of $H$ with energy $E$, we compute the portion of its total weight that lies within the vector space
\begin{align}\label{eq:Vpi}
\mathcal V_\pi 
=
\text{span}
\left\{
|n\rangle,|\tilde n\rangle
\mid
n=0,\dots,L/2
\right\}.
\end{align}
$\mathcal V_\pi$ contains a total of $L+2$ states, half of them obtained from successive applications of $S^+_\pi$ to the ground state and the other half from successive applications of $S^-_\pi$ to the ceiling state. (Note that the system size in our study is always even.)  
More concretely, denoting by $|e_i\rangle$ ($i=1,\dots,L+2$) an orthonormal basis for $\mathcal V_\pi$ (which can be found using a standard Gram-Schmidt orthogonalization procedure), we compute the total weight
\begin{align}\label{eq:weight}
W_{\mathcal V_{\pi}}(E)=\sum^{\text{dim } \mathcal V_\pi}_{i=1}|\langle e_i|E\rangle|^2 \leq 1.
\end{align}
The weight $W_{\mathcal V_{\pi}}(E)$ attains the maximal value $1$ when the state $|E\rangle$ is precisely a linear combination of the $|e_i\rangle$.  For a typical eigenstate in the middle of the many-body spectrum, this weight is exponentially small, i.e.~$W_{\mathcal V_{\pi}}(E)\sim \text{dim }\mathcal V_\pi/\mathcal D\sim L/\varphi^L$, where $\mathcal D$ is the Hilbert-space dimension.

%%%%%%%%%%%%
%%%%%%%%%%%%
\begin{figure}[t!]
\begin{center}
\includegraphics[width=0.946\columnwidth]{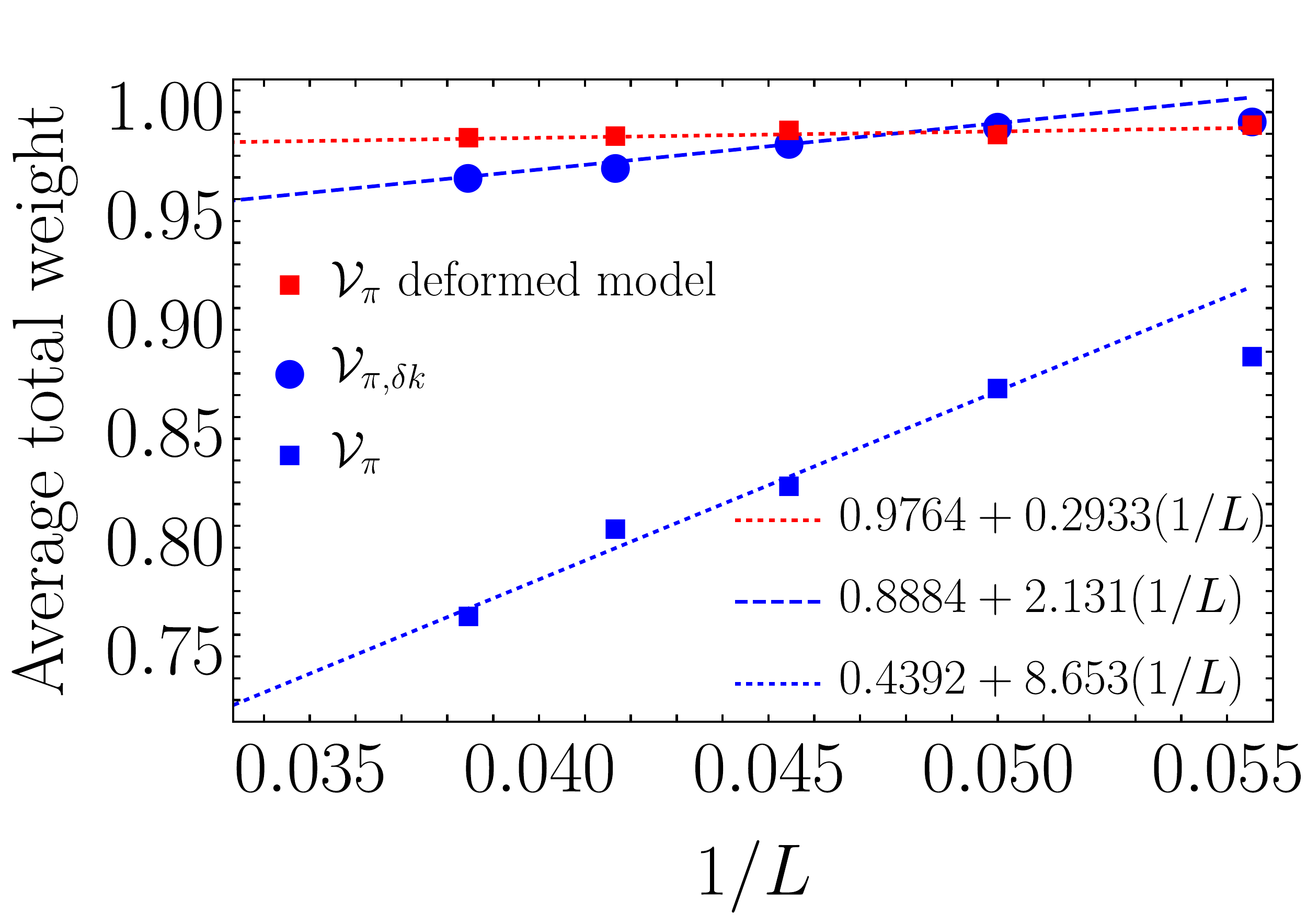}
\caption{
System-size dependence of the average of the total weight \eqref{eq:weight} over the set of primary scarred eigenstates at $k=0$.  The average total weights on the vector spaces $\mathcal V_\pi$ (blue squares) and $\mathcal V_{\pi,\delta k}$ (blue dots) of the primary scar states in the model \eqref{eq:H} are plotted as a function of $1/L$ for $L=18,\dots,26$. The average total weight on the vector space $\mathcal V_\pi$ for the primary scar states in the deformed model \eqref{eq:Hd} (red squares) is plotted for the same values of $L$.  Dashed and dotted lines are linear extrapolations~\cite{Note2},
listed in the inset, whose $y$-intercepts correspond to predictions for the average total weights in the thermodynamic limit. For the model \eqref{eq:H}, a clear improvement results from using the vector space $\mathcal V_{\pi,\delta k}$ instead of $\mathcal V_\pi$. For the model \eqref{eq:Hd}, the vector
space $\mathcal V_\pi$ is sufficient to account for the overwhelming majority of the weight of a typical scar state.
}
\label{fig:tower-basis}
\end{center}
\end{figure}
%%%%%%%%%%%%
%%%%%%%%%%%%

In Fig.~\ref{fig:tower-basis}, we show the total weight $W_{\mathcal V_{\pi}}(E)$ averaged over the set of primary scar states in the zero-momentum sector for system sizes $L=18,\dots,26$.  The average total weight shows a decreasing trend with increasing system size, from $\sim 0.89$ at $L=18$ to $\sim 0.77$ at $L=26$. For comparison, the infinite-temperature average of $W_{\mathcal V_\pi}(E)$ over the zero-momentum sector at $L=26$ gives $\text{dim }\mathcal V_\pi/\mathcal D=0.002676\dots$.  A linear extrapolation to 
$L=\infty$~\footnote{Although the data presented here fit reasonably well to linear functions for the range of system sizes considered, another functional form that could be used for extrapolation is $A\, e^{-\varepsilon L}$ with fitting parameters $A$ and $\varepsilon$. This exponential form could arise if the variational basis being used makes a small but finite error $\sim \varepsilon$ per unit volume.  Our data also fit well this functional form for the system sizes considered.  For this exponential extrapolation, we find that our data fit to an exponential with $\varepsilon = 0.01814$ for $\mathcal V_\pi$, $\varepsilon=0.00360$ for $\mathcal V_{\pi,\delta k}$, and $\varepsilon=0.00062$ for $\mathcal V_\pi$ in the deformed model of Sec.~\ref{sec:deform}.  These numbers should be compared with $\ln\varphi=0.48121$, which would arise for a random state in the constrained Hilbert space. Thus, our data still compare favorably with the generic case if one uses exponential rather than linear extrapolation.} 
yields an average total weight of $0.4392$, suggesting that a typical scar state has finite overlap with the $\pi$-magnon tower of states spanning $\mathcal V_\pi$ in the thermodynamic limit. 

\subsection{Relative Momentum and $\pi$-Magnon Scattering}\label{sec:tower_b}

The $\pi$-magnon description of the primary scar states can be improved systematically by taking into account the existence of multimagnon interactions.  For example, in principle nothing prevents two magnons with initial momentum $\pi$ from scattering off one another such that the two magnons have final momenta $\pi+\delta k$ and $\pi-\delta k$, respectively. Even in the absence of a microscopic understanding of such processes, one can account for them by defining multimagnon states in which pairs of magnons are allowed to have nontrivial relative momenta.

For example, we can define the $n$-magnon states ($n\geq 2$)
\begin{align}\label{eq:ndkstates}
\begin{split}
|n,\delta k\rangle &= \mathcal N_{n,\delta k}\,
(S^+_{\pi})^{n-2}
\\
&\qquad
\times
(S^+_{\pi+\delta k}S^+_{\pi-\delta k}
\!+\!
S^+_{\pi-\delta k}S^+_{\pi+\delta k})
|{\rm GS}\rangle,
\end{split}
\end{align}
where one pair of magnons has nontrivial relative momentum $\delta k = 0,2\pi/L,\dots,\pi$ while the remaining $n-2$ all have momentum $\pi$.  Note that the operator in parentheses above, which creates the pair of magnons with nontrivial relative momentum, is symmetrized with respect to exchanging the momenta of the pair in order to be consistent with the bosonic statistics of magnons.  By analogy with Eq.~\eqref{eq:ntilstates}, we also define the reflected states
\begin{align}\label{eq:ntildkstates}
|\tilde n,\delta k\rangle= C\, |n,\delta k\rangle.
\end{align}
Note that the states \eqref{eq:ndkstates} and \eqref{eq:ntildkstates} reduce to the states \eqref{eq:nstates} and \eqref{eq:ntilstates} when $\delta k=0$.

To determine the extent to which the states \eqref{eq:ndkstates} and \eqref{eq:ntildkstates} improve the description of the scar states, we proceed as we did for the states \eqref{eq:nstates} and \eqref{eq:ntilstates}. By analogy to Eq.~\eqref{eq:Vpi}, we define the vector space
\begin{align}\label{eq:Vpi,dk}
\mathcal V_{\pi,\delta k}
&=
\text{span}
\Bigg\{
\left\{
|0\rangle,
|\tilde 0\rangle,
|1\rangle,
|\tilde 1\rangle
\right\},\\
&
\left\{
|n,\delta k\rangle,|\tilde n,\delta k\rangle
\mid
n=2,\dots,\frac{L}{2};
\delta k=0,\dots,\pi
\right\}
\Bigg\},
\nonumber
\end{align}
a basis for which is again obtained by Gram-Schmidt orthogonalization. The dimension of $\mathcal V_{\pi,\delta k}$ now scales as $L^2$, as opposed to $\mathcal V_{\pi}$, whose dimension scales as $L$.  We subsequently compute the weight $W_{\mathcal V_{\pi,\delta k}}(E)$, defined by analogy with Eq.~\eqref{eq:weight}, for each scar state on the vector space $\mathcal V_{\pi,\delta k}$. 

The average of this weight over all scar states at $k=0$ is plotted as a function of system size in Fig.~\ref{fig:tower-basis}.  As was the case for the $\pi$-magnon vector space $\mathcal V_\pi$, the average total weight decreases as a function of system size.  However, the numerical value of the average weight at fixed system size increases substantially with the new choice of basis, ranging from $\sim0.997$ at $L=18$ to $\sim 0.97$ at $L=26$. (For comparison, the infinite-temperature result in the zero-momentum sector at $L=26$ is $ \text{dim }\mathcal V_{\pi,\delta k}/\mathcal D=0.01625\dots$.)  Moreover, the estimated average weight at $L=\infty$ based on linear extrapolation doubles, increasing from $0.4392$ to $0.8884$. 

Remarkably, this improvement results from including states in which only \textit{one pair} of magnons has nontrivial relative momentum.  The procedure outlined above can in principle be repeated \textit{ad nauseam} by constructing vector spaces from states $|n,\delta k_1,\delta k_2,\dots\rangle$ in which additional pairs of magnons are allowed to have nontrivial relative momentum.  The natural termination of this procedure arises when \textit{every} pair of magnons is allowed to have an independent relative momentum, resulting in a basis whose size scales exponentially with system size.  Evidently, such a description is not necessary in order to describe the scar states with high accuracy--a polynomially large basis suffices.  This observation suggests the possibility that magnons with momentum $\pi$ scatter more weakly off of one another than magnons at arbitrary momenta, which could provide a mechanism whereby certain many-body eigenstates can be well described as states with a macroscopic number of magnons at momentum $\pi$.  A more thorough investigation of magnon scattering in the PXP model would thus be an interesting avenue for future work.

It is interesting to note that the vector space $\mathcal V_{\pi,\delta k}$ also captures certain aspects of the ``descendant" (i.e., non-primary) scar states visible in Fig.~\ref{fig:overlap}.  To demonstrate this, we consider the projection $\mathcal P_{\mathcal V_{\pi,\delta k}}|E\rangle$ of every eigenstate $|E\rangle$ onto $\mathcal V_{\pi,\delta k}$. Although generic eigenstates indeed have exponentially small weight on $\mathcal V_{\pi,\delta k}$, there are a handful of eigenstates with substantial weight on $\mathcal V_{\pi,\delta k}$.  For example, at $L=24$ and $k=0$ there are $53$ eigenstates whose weight on $\mathcal V_{\pi,\delta k}$ exceeds $0.6$.  In Fig.~\ref{fig:overlap}, we plot the overlap with the N\'eel state and the energy expectation value of the (normalized) state $\mathcal P_{\mathcal V_{\pi,\delta k}}|E\rangle/\sqrt{W_{\mathcal V_{\pi,\delta k}}(E)}$ for eigenstates $|E\rangle$ with $W_{\mathcal V_{\pi,\delta k}}(E)>0.6$.  Comparing with the exact overlaps and energies, there is clear qualitative (and in some cases, including those of the primary scar states, quantitative) agreement between the projected and exact states.

%%%%%%%%%%%%
%%%%%%%%%%%%
\begin{figure}[t!]
\begin{center}
\includegraphics[width=0.85\columnwidth]{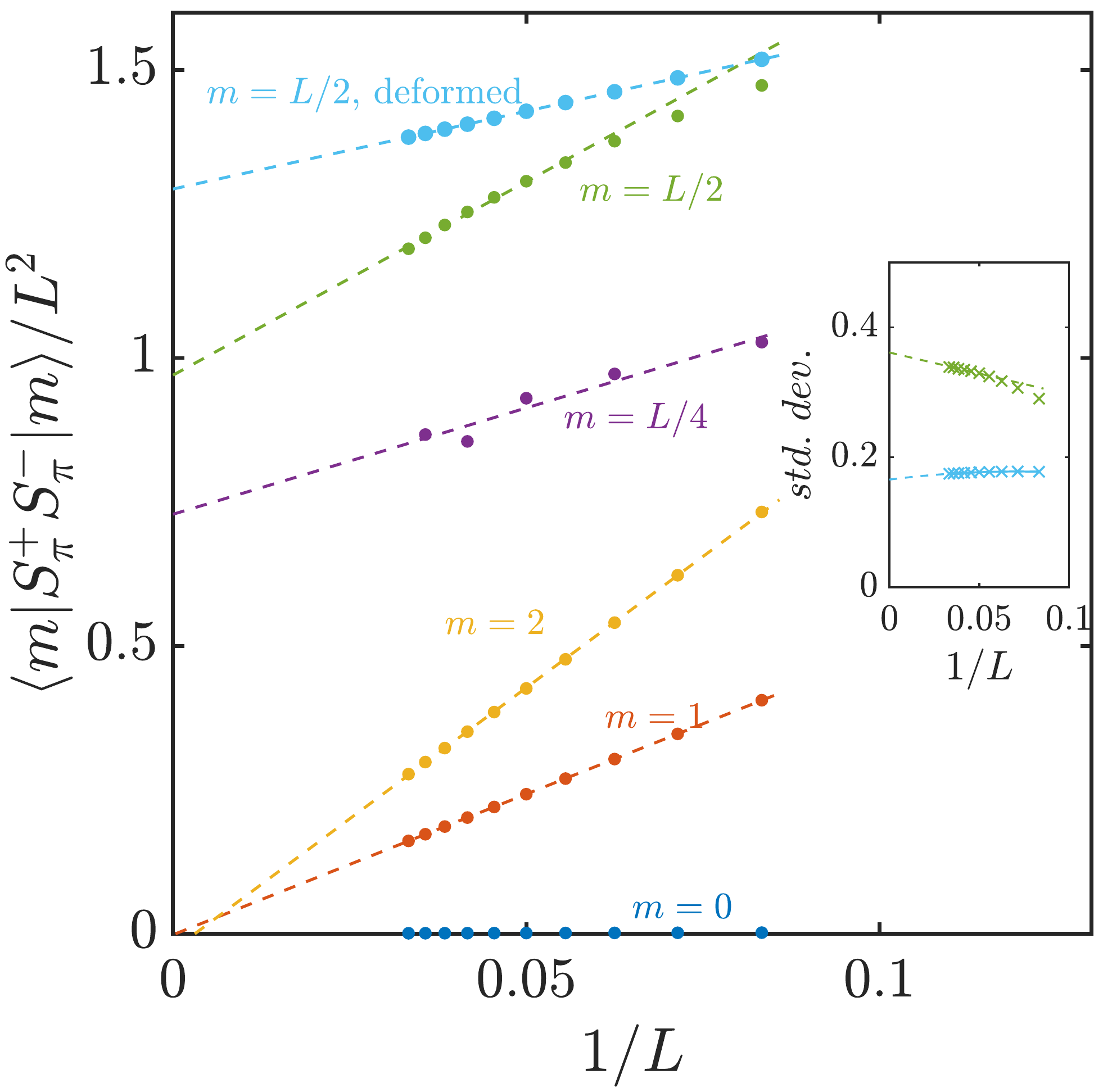}
\caption{$\pi$-magnon density evaluated in the primary scar states, labeled by the primary scar index $m=0$ (ground state)$,\dots,L$ (ceiling state), as a function of $1/L$ for $L=12,\dots,30$. For scar states at finite energy density (i.e.,~finite $m/L$) the $\pi$-magnon density extrapolates to a finite value, indicating macroscopic $\pi$-magnon occupation. Furthermore, the magnon number fluctuations (inset) are also macroscopic, indicating the presence of magnon condensation. The light blue data points are results from the deformed Hamiltonian, Eq.~\eqref{eq:Hd}, which show enhanced $\pi$-magnon order and reduced, yet still macroscopic, number fluctuations (inset).}
\label{fig:order1}
\end{center}
\end{figure}
%%%%%%%%%%%%
%%%%%%%%%%%%

\section{Consequences of the $\pi$-Magnon Description: Spatiotemporal Long-Range Order}\label{sec:order}

Having established in Sec.~\ref{sec:tower} that the scar states at finite energy density can be well-described using states containing macroscopic numbers of predominantly $\pi$-momentum magnons, we now discuss several interesting implications that derive from this fact. 

\subsection{Scar States as $\pi$-Magnon Condensates}\label{sec:condensate}

Since magnons are bosonic quasiparticles, it is natural to anticipate that the scar states can be viewed as Bose-Einstein condensates of $\pi$-magnons. A scar state can be viewed as a $\pi$-magnon condensate if it possesses off-diagonal long-range order (ODLRO)~\cite{Yang62} associated with the $\pi$-momentum component of the one-magnon reduced density matrix, i.e.,
\begin{equation}\label{eq:order}
\langle S^+_\pi S^-_\pi \rangle/L^2 \to {\rm const.} \text{ as } L\to\infty.
\end{equation}
The criterion \eqref{eq:order} is equivalent to the statement that the state contains a finite density of $\pi$-magnons in the thermodynamic limit.
In Fig.~\ref{fig:order1} we show numerically via finite-size extrapolation to $L=\infty$ of data from $L=12,\dots,30$ that this is indeed the case for scar states at finite energy density (i.e., those possessing a macroscopic number of $\pi$-magnons).

It is important to note that in systems described by the canonical ensemble, Bose-Einstein condensation is associated with the spontaneous breaking of a U$(1)$ particle-number symmetry. It is evident from Eq.~\eqref{eq:H} that no such symmetry exists in the PXP model. However, it is well-known that Bose-Einstein condensation may also be described using the grand canonical ensemble, where particles may be exchanged with an external reservoir. In this case, the phase-coherent condensate exhibits \emph{macroscopic} number fluctuations. The role of the reservoir is crucial in this description (even though the coupling is weak) because it defines the symmetry-breaking field to which the system is infinitely susceptible. We show in the inset of Fig.~\ref{fig:order1} that the $\pi$-magnon number fluctuations are also macroscopic for scar states at finite energy density.

This leads to an intriguing possibility: that $\pi$-magnon condensation may be loosely understood by an analogy to the grand canonical ensemble wherein the $\pi$-magnons define a bosonic system weakly coupled to an effective magnon reservoir (e.g. due to thermal states in the spectrum). Determining the extent to which this analogy holds will require understanding how ``isolated'' the $\pi$-magnon sector actually is in the PXP model and the microscopic mechanism behind it. Such questions are interesting topics for future investigations.

\subsection{Space-Time Crystalline Order in Scar States}\label{sec:space-time}

%%%%%%%%%%%%
%%%%%%%%%%%%
\begin{figure}[t!]
\begin{center}
\includegraphics[width=\columnwidth]{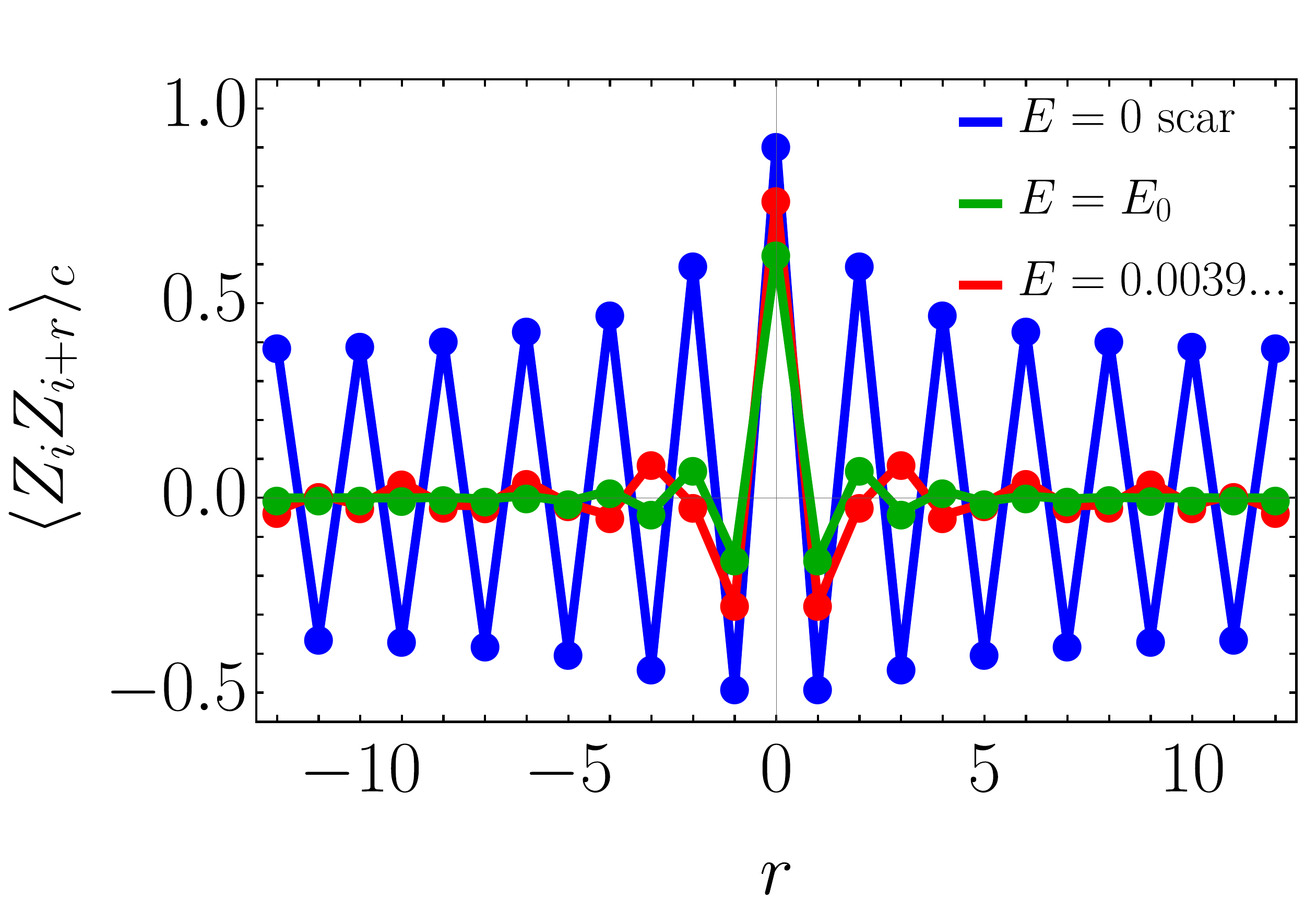}
\caption{Connected spin correlations in real space at $L=24$ for several eigenstates. Scar states at finite energy density (e.g. the $E=0$ scar state) exhibit long-range crystalline order at wavevector $k=\pi$ associated with $\pi$-magnon condensation, see Eq.~\eqref{eq:lro}. In a typical state in the middle of spectrum (near $E=0$) the correlations decay as $r\to \infty$, consistent with the ETH prediction, Eq.~\eqref{eq:eth-order}.}
\label{fig:order2}
\end{center}
\end{figure}
%%%%%%%%%%%%
%%%%%%%%%%%%

\subsubsection{Spatial Eigenstate Ordering}

An important consequence of $\pi$-magnon condensation in the primary scar states relates to the structure of real-space correlations in these states. Indeed, if a macroscopic number of excitations exist at a finite momentum $k=\pi$ in such states, then they must exhibit long-range order (LRO) in the form of a ``magnon density wave" with wavenumber $\pi$. This implies that operators sensitive to the magnon number will exhibit spatially oscillating correlations with a two-site period in the primary scar states. In particular, since $Z_i$ is utilized in the construction of the magnon creation and annihilation operators, c.f.~Eq.~\eqref{eq:s+-}, the scar states will exhibit long-range connected correlations of the form
\begin{equation}\label{eq:lro}
\langle Z_i Z_{i+r}\rangle_{c,\,{\rm scar}}\sim(-1)^r\times {\rm const.},\,\,\,r\to\infty,
\end{equation} 
where the subscript $c$ denotes connected correlations and the prefactor depends on the energy of the scar state, being finite only for states with finite energy density (or, equivalently, finite $\pi$-magnon density), consistent with Fig.~\ref{fig:order1}.

In Fig.~\ref{fig:order2} we plot the spin correlations along the $z$-direction for a selection of representative states (correlations along the $y$-direction show a similar behavior). We find that the scar states at finite energy density exhibit oscillations at wavenumber $k=\pi$ that do not decay with increasing spatial separation $r$, while a typical state (or scar states at zero energy density) do decay as $r\to\infty$, cf.~Fig.~\ref{fig:order2}. This provides further evidence for the breakdown of the strong ETH, which predicts that infinite temperature correlations decay as 
\begin{equation}\label{eq:eth-order}
\langle Z_i Z_{i+r}\rangle_{c,\,{\rm ETH}} \sim (-1)^r \varphi^{-2r},\,\,\,r\to\infty ,
\end{equation}
where $\varphi$ is the golden ratio. While this behavior is consistent with typical states in the middle of the many-body spectrum (as shown in Fig.~\ref{fig:order2}), it is not true of the scarred states which instead show correlations of the form~\eqref{eq:lro}.

\subsubsection{Temporal Eigenstate Ordering}

Another remarkable consequence of $\pi$-magnon condensation in the primary scar states has to do with their \emph{dynamical} correlations: the primary scar states exhibit long-range order in \emph{time} as well as space. In particular, as we now argue, the primary scar states near the middle of the many-body spectrum exhibit unequal-time correlations of the form
\begin{equation}\label{eq:dynamics}
\langle Z_\pi(t) Z_\pi(0)\rangle_{\rm scar}/L^2\to f(t) \text{ as } L\to\infty,
\end{equation}
where $f(t)$ is some periodic function of time with period $2\pi/\Omega$ and $\Omega\approx 1.33$ is the energy spacing between scar states near the middle of the spectrum.  Eq.~\eqref{eq:dynamics} is the definitive~\cite{Watanabe15} signature of space-time crystalline order (see Ref.~\cite{Sacha17} for a review), which we argue below is  associated with $\pi$-magnon condensation. We note that such order does not contradict the established no-go theorems~\cite{Bruno13,Watanabe15} that preclude time-crystal order in ground states and at thermal equilibrium; the scar states are at finite energy density, do not obey ETH, and represent a vanishing fraction of all eigenstates. Thus, the space-time crystalline order in the scar states is a purely nonequilibrium phenomenon. 

%%%%%%%%%%%%
%%%%%%%%%%%%
\begin{figure}[t!]
\begin{center}
\includegraphics[width=0.9\columnwidth]{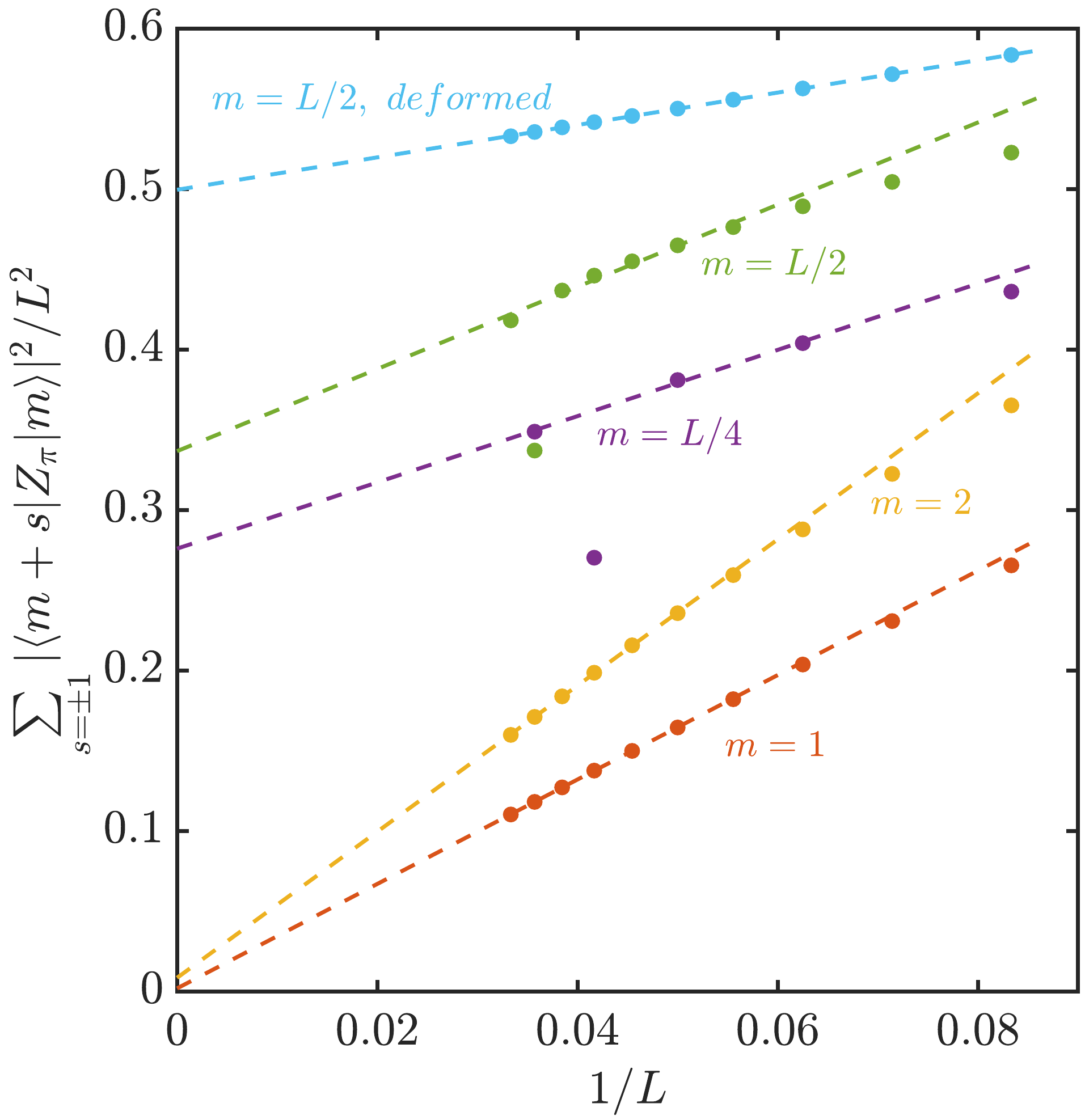}
\caption{Scaling of the matrix elements of $Z_\pi$ entering Eq.~\eqref{eq:dynamics2} as a function of system size ($L=12,\dots,30$) for primary scar states with index $m$ (defined as in Fig.~\ref{fig:order1}).  The $Z_\pi$ matrix elements for scar states near the ground state ($m=1,2$) evidently scale to zero as $1/L$, similar to what is observed for the magnon density in Fig.~\ref{fig:order1}.  Matrix elements for scar states at finite energy density ($m=L/4,L/2$) appear to scale towards a constant value as $L$ increases, a necessary condition for long-range order in time.  The outliers visible at certain system sizes are the result of weak hybridization with generic eigenstates close by in energy, similar to what was observed for the entanglement entropy in Ref.~\cite{Turner18}.  The light blue data points correspond to the deformed model discussed in Sec.~\ref{sec:deform}, where the $Z_\pi$ matrix elements are enhanced and the outliers no longer present.  These results are consistent with those presented in Fig.~\ref{fig:order1}. }
\label{fig:Z_pi}
\end{center}
\end{figure}
%%%%%%%%%%%%
%%%%%%%%%%%%

To see how such temporal long-range order can arise, we re-express the correlation function on the left-hand side of \eqref{eq:dynamics} as
\begin{align}\label{eq:dynamics2}
\langle m|Z_\pi(t) Z_\pi(0)|m\rangle &= e^{-\mathrm{i}\Omega t}\, |\langle m+1|Z_\pi|m\rangle|^2\\
&\qquad + e^{+\mathrm{i}\Omega t}\, |\langle m-1|Z_\pi|m\rangle|^2+\dots,\nonumber
\end{align}
where $|m\rangle$ ($m=0,\dots, L$) is a primary scar state and $\dots$ denotes terms depending on matrix elements of $Z_\pi$ between the state $|m\rangle$ and all other eigenstates.  From Eq.~\eqref{eq:dynamics2}, it is clear that the condition \eqref{eq:dynamics} holds if
\begin{align}\label{eq:dynamics3}
\langle m\pm1|Z_\pi|m\rangle/L\to\text{const.}
\end{align}
as $L\to\infty$, while all other off-diagonal matrix elements
$\langle E|Z_\pi|m\rangle\to 0$~\cite{foot2}.
The ETH~\cite{D'Alessio16} predicts that generic off-diagonal matrix elements of local operators are exponentially small in system size---thus, Eq.~\eqref{eq:dynamics3} is only possible when the states $|m\rangle$ and $|m\pm 1\rangle$ are ETH-violating.

%%%%%%%%%%%%
%%%%%%%%%%%%
\begin{figure}[t!]
\begin{center}
\includegraphics[width=.99\columnwidth]{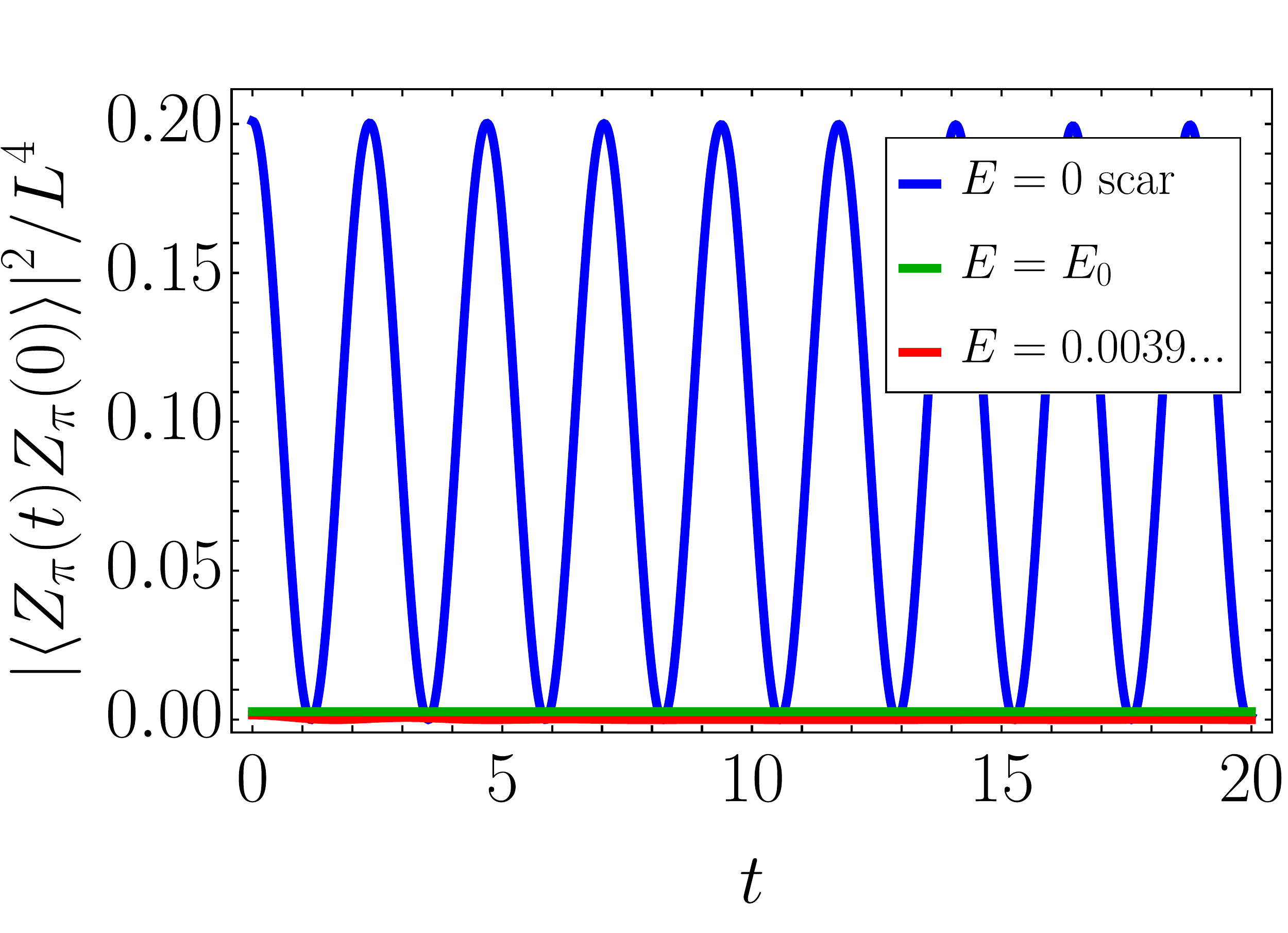}
\caption{Unequal-time spin correlations at momentum $\pi$ in eigenstates of Eq.~\eqref{eq:H}. (Note that only the connected part contributes due to the fact that we are considering finite-wavenumber correlations.) Spin correlations in scar states at finite energy density show persistent oscillations, consistent with Eq.~\eqref{eq:dynamics}. The frequency of the oscillations is set by the effective $\pi$-magnon chemical potential, which separates neighboring scar states in energy (the energy spacing is $\Omega\approx1.33$ at $L=24$). For generic eigenstates the oscillations decay (and in fact have a prefactor scaling to zero as $L\to\infty$) due to the absence of magnon condensation.}
\label{fig:dynamics}
\end{center}
\end{figure}
%%%%%%%%%%%%
%%%%%%%%%%%%

Intuitively, the scaling in Eq.~\eqref{eq:dynamics3} is plausible within the magnon-condensate picture of the scar states after recalling from Eq.~\eqref{eq:s+-} that $Z_\pi=S^+_\pi+S^-_\pi$.  Indeed, as shown in Fig.~\ref{fig:order1}, the number of $\pi$-magnons in the scar states $|m\rangle$ increases with increasing $m$ and becomes macroscopic.  Thus, it is reasonable to expect that matrix elements of $Z_\pi$ between consecutive scar states $|m\rangle$ and $|m\pm1\rangle$ would be much larger than matrix elements between the state $|m\rangle$ and other eigenstates of $H$.  We provide numerical evidence in Fig.~\ref{fig:Z_pi} that Eq.~\eqref{eq:dynamics3} holds for primary scar states $|m\rangle$ with $m\sim L$.  In Fig.~\ref{fig:dynamics}, we demonstrate that the hypothesized space-time-periodic connected correlations arise in a scar state in the middle of the many-body spectrum for $L=24$, and contrast with the cases of the ground state and a generic finite-energy density eigenstate, which do not show such correlations.

The relationship between $\pi$-magnon condensation and space-time crystalline order is in fact rather natural.  Indeed, it was noted early on that condensates at finite chemical potential can yield examples of time crystals~\cite{Wilczek13}, provided the initial state is not an equilibrium state~\cite{Watanabe15,Else17}, with the caveat that the Hamiltonian must contain U$(1)$-breaking terms in order for the oscillations to be observable~\cite{Volovik13,Watanabe15,Else17}.  Such terms could arise due to, e.g., coupling the condensate to a bath~\cite{Volovik13} or another condensate~\cite{Watanabe15}.  This scenario arises naturally in the context of the scarred eigenstates of the PXP model: these eigenstates contain macroscopic numbers of $\pi$-magnons and thus exhibit the ODLRO of a condensate, but the magnon number is not conserved.   

It is interesting to interpret the coherent dynamics observed experimentally in Ref.~\cite{Bernien17} in the language of time-crystalline order.  While the authors of Ref.~\cite{Bernien17} did not measure connected correlations in eigenstates of Eq.~\eqref{eq:H} as we do in Fig.~\ref{fig:dynamics}, they prepared an initial state, the N\'eel state, that has disproportionately high overlap with the scarred eigenstates and thus effectively projects the dynamics onto these states.  Because this initial state happens to be a $z$-basis product state, the connected part of the spatial correlator $\langle Z_i(t)Z_{i+r}(0) \rangle$ vanishes. However, in Fig.~6a of Ref.~\cite{Bernien17} it is clear that there is a spatially oscillatory component to the measured correlation function in addition to the temporal oscillations that motivated subsequent work on quantum many-body scars.  The fact that the system maintains spatial as well as temporal coherence during the evolution can be viewed as an interaction effect associated with the long-range order of the scarred eigenstates in both space and time.  This idea can be further tested experimentally by measuring the correlation function $\langle Y_{i}(t)Y_{i+r}(0)\rangle$, again starting from the N\'eel state.  The connected part of this correlator also exhibits nontrivial spatiotemporal oscillations in the scarred eigenstates at finite energy density, but now the initial state is no longer an eigenstate of the operator being measured, so that a nontrivial connected part can be extracted from dynamics.

\section{Model Deformation}\label{sec:deform}

It is intriguing to consider whether the eigenstate order detailed in Sec.~\ref{sec:order} can be considered as a distinct nonequilibrium or ``eigenstate" phase~\cite{Parameswaran17}. For this to be the case, such order would need to be robust with respect to small changes to the Hamiltonian~\eqref{eq:H}. While we will not consider here an exhaustive study of model deformations, we will show below that a deformation studied in Refs.~\cite{Choi18,Khemani18} leads to an \emph{enhancement} of the $\pi$-magnon description of the scar states and the associated LRO.  This deformation was shown in Ref.~\cite{Choi18,Khemani18} to dramatically enhance the coherence and lifetime of the oscillations due to the scarred eigenstates.  That the same deformation also enhances the $\pi$-magnon description and LRO introduced in this work further emphasizes that these aspects are true hallmarks of the scarred eigenstates and should be considered essential features of the phenomenon.

We consider the deformed Hamiltonian $H^\prime=H+\delta H_R$, where
\begin{equation}\label{eq:Hd}
\delta H_R=-\sum_i \sum_{d=2}^R h_d\, P_{i-1}X_i P_{i+1}\left(Z_{i-d}+Z_{i+d}\right)
\end{equation}
was introduced in Refs.~\cite{Choi18,Khemani18}.  Here $R$ denotes the range of the deformation. The deformation $\delta H_2$ was introduced in Ref.~\cite{Khemani18}, where it was found to both enhance the oscillations and reduce level repulsion in the energy spectrum for a certain deformation strength, suggesting the possibility of proximity to an as-yet-unknown integrable point.
In Ref.~\cite{Choi18}, $R>2$ was considered and the parameters $h_d$ were tuned to optimize the fidelity of the revivals after the quench from the N\'eel state. It was shown that this can be accomplished using
\begin{equation}\label{eq:hd}
h_d=h_0\left(\varphi^{d-1}-\varphi^{-(d-1)}\right)^{-2},
\end{equation}
where $h_0\approx 0.051$ is obtained by optimizing the FSA.  Studies of energy level statistics for $H^\prime$~\cite{Choi18} indicate that the deformation with parameters \eqref{eq:hd} does not drive the system closer to the putative integrable point of Ref.~\cite{Khemani18}.

We find that adding the deformation~\eqref{eq:Hd} up to range $R=10$ significantly improves the $\pi$-magnon description of the scar states. In particular, the extrapolated value of the average total weight of the primary scar states on the $\pi$-magnon vector space $\mathcal V_\pi$ defined in Eq.~\eqref{eq:Vpi} increases from $0.4392$ to $0.9764$, see Fig.~\ref{fig:tower-basis}. Remarkably, this enhancement of the average weight occurs using the \emph{same} magnon ladder operators defined in Eq.~\eqref{eq:s+-} with $\alpha=2$. Moreover, the relative enhancement of the weight obtained by replacing $\mathcal V_\pi$ with the enlarged vector space $\mathcal V_{\pi,\delta k}$, defined in Eq.~\eqref{eq:Vpi,dk}, diminishes for the deformed model. Whereas for the model~\eqref{eq:H} the extrapolated average total weight more than doubles from $0.4392$ to $0.8884$ (see Fig.~\ref{fig:tower-basis}), for the deformed model the extrapolated averaged weight merely increases from $0.9764$ to $0.9970$. This indicates that the pure $\pi$-magnon vector space $\mathcal V_\pi$ provides a much more accurate description of the scar states in the deformed model than it does in the original PXP model~\eqref{eq:H}, suggesting that the $\pi$-magnons become further isolated and interact more weakly with one another in the deformed model. 

In addition, the $\pi$-magnon condensation and associated LRO are enhanced by the deformation~\eqref{eq:Hd}. For example, in the $E=0$ scar state the extrapolated value of $\langle S^+_\pi S^-_\pi\rangle/L^2$ increases from $0.98$ to $1.33$, while the number fluctuations are reduced (but remain macroscopic), see Fig.~\ref{fig:order1}. These results for the deformed model thus suggest that the phenomena reported here are robust, at least up to the system sizes ($L\sim 30$) accessible by exact diagonalization.

\section{Discussion and Conclusion}\label{sec:conclusion}
In this paper we explored the properties of atypical ``scarred" eigenstates of the PXP model, Eq.~\eqref{eq:H}, which was recently realized in a Rydberg-atom quantum simulator. We showed that these special states can be accurately represented by a basis of states containing macroscopic numbers of magnons with momentum $\pi$ (see Sec.~\ref{sec:tower}).  This description is surprising given that generic scarred eigenstates have finite energy density, while $\pi$-magnons are the lowest-lying excitations above the ground state (see Sec.~\ref{sec:sma}); in generic nonintegrable quantum many-body systems such a quasiparticle description is not operative at finite energy density. The $\pi$-magnon description of the scar states leads to a remarkable set of new predictions for these states at finite energy density, which we test numerically in Sec.~\ref{sec:order}. In particular, we demonstrated that the finite-energy-density scar states exhibit long-range order associated with $\pi$-magnon condensation. The condensation of magnons at momentum $\pi$ combined with the near-constant energy spacing between scar states in the middle of the many-body spectrum gives rise to persistent density oscillations in both space and time. Thus, the scarred states can be regarded as having nonequilibrium space-time crystalline order. We also demonstrated that a deformation~\cite{Choi18,Khemani18} of the PXP model that enhances the fidelity of the post-quench revivals also enhances the $\pi$-magnon description and long-range order that we uncovered in this work.  This suggests that the phenomena we discuss are both stable and essential properties of the dynamical ``phase" defined by the emergence of scarred eigenstates (to the extent that such a phase exists).  An interesting test of this idea for future work would be to examine whether a picture similar to the one developed here can also be applied to generalizations of the PXP model like those studied in Refs.~\cite{Ho19,Bull19}, which also appear to support scarred eigenstates.

It is important to note that the quasiparticle picture developed in this work is very different from the one suggested in Ref.~\cite{Lin18} and extended in Ref.~\cite{Surace19}.  In those works, the finite-energy-density scar states near zero energy (i.e., in the middle of the many-body spectrum) are approximated by an SMA-like variational Ansatz on top of the exact zero-energy eigenstate found in Ref.~\cite{Lin18}. (Interestingly, this exact eigenstate is not a primary scar state, i.e., there are many other zero-energy eigenstates~\cite{Turner17,Schecter18}, some of which have higher overlap with the N\'eel state.)  In contrast, the $\pi$-magnon description of the scar states developed here involves elementary excitations above the ground state.  These two quasiparticle pictures lead to very different predictions for the scaling of the scar-state entanglement entropy with system size in the thermodynamic limit: while the $\pi$-magnon picture in this work is naturally consistent with the logarithmic entanglement scaling found in Ref.~\cite{Turner18}, the quasiparticle picture of Refs.~\cite{Lin18,Surace19} is not. This is because the variational Ansatz states used in Refs.~\cite{Lin18,Surace19} to approximate the scar states closest to zero energy arise from applying a local operator to a finite-bond-dimension MPS, and as such must have area-law entanglement.  It is also unclear whether the quasiparticle picture of Refs.~\cite{Lin18,Surace19} can be used to explain the eigenstate properties uncovered in this work.  A more detailed study of the quasiparticle picture of Refs.~\cite{Lin18,Surace19} would be necessary to clarify these points.

In the future it will be important to understand the microscopic mechanism behind the stability of the $\pi$-magnon condensation reported here, as well as why these excitations remain ``well-isolated'' from magnon states with other momenta. One possible explanation discussed in Sec.~\ref{sec:su(2)} and in Ref.~\cite{Choi18} is the emergence of an effective SU$(2)$ algebra (similar to $\eta-$pairing in the Hubbard model) where the true $\pi$-magnon creation operator becomes an eigenoperator of the Hamiltonian. If such an algebra does indeed emerge in some point or region of parameter space, it would be interesting to track the properties of the $\pi$-magnons and their scattering as this point/region is approached. Our work suggests the possibility of shedding light on these questions using well-established matrix product state methods for extracting scattering properties of low-lying excitations in spin chains~\cite{Vanderstraeten14}.  Studying how $\pi$-magnons interact with magnons at other momenta at low energies could allow one to develop a ``bootstrapped" understanding of the finite-energy density scar states; indeed, the central message of this paper is that low-lying excitations and their properties help to determine the characteristics of these special highly-excited eigenstates.

More broadly, much remains to be understood about the presence of quantum many-body scars in the PXP model and its relatives.  While the stability of the coherent dynamics and associated eigenstate properties to perturbations has already been discussed here and in Refs.~\cite{Choi18,Khemani18}, the basic question of the existence of these properties for the PXP model in the thermodynamic limit is not yet settled.  The lack of progress on this front is due to the strongly interacting nature of the problem and the limited system sizes accessible to exact diagonalization, for which finite-size extrapolation may be misleading.  Hints that problems may appear as the thermodynamic limit is taken have already arisen in existing data.  For example, the entanglement entropy scaling in Fig.~8 of Ref.~\cite{Turner18} exhibits non-monotonic behavior that can be attributed to hybridization between the scar states and other volume-law states nearby in energy.  Another example is the outliers in our Fig.~\ref{fig:Z_pi} at $m=L/4,L/2$, which can also be attributed to such hybridization.  It remains to be seen whether such resonances ultimately destroy the scar states and make them generic as $L\to\infty$.  However, the fact that there exists a deformation that enhances the atypical properties of the scar states, including the long-range order studied in this work, seems to suggest that there may exist a point or small region in parameter space where such properties are truly stable in the thermodynamic limit.  Further progress on this question will likely require an improved analytic understanding of the scarred eigenstates, and we anticipate that the $\pi$-magnon picture developed in this work will contribute to further progress in this direction.

Finally, it is interesting to consider other contexts in which quantum many-body scars can arise, beyond the examples of $\eta$-pairing states and bimagnons in the AKLT chain that we have already discussed.  For instance, Ref.~\cite{Surace19} suggests that phenomena analogous to the long-lived oscillations observed in the PXP model can also arise in lattice gauge theories, which can show slow dynamics due to a mechanism reminiscent of the Schwinger effect.  (In fact, the authors provide a mapping between the PXP model and such a gauge theory.  It would be interesting to reinterpret the results of our work in the lattice gauge theory context.). In Ref.~\cite{Iadecola18}, an exact family of finite-energy-density eigenstates with area-law entanglement was found in a class of otherwise ETH-obeying spin ladders.  More generally, Refs.~\cite{Shiraishi17} and \cite{Ok19} have proposed systematic constructions of Hamiltonians with exact quantum many-body scar states using local projection techniques, and Ref.~\cite{Pai19} has proposed a distinct mechanism related to symmetry constraints on quantum dynamics.  Furthermore, Ref.~\cite{Freivogel12} found that persistent oscillations after quantum quenches from generic initial states can arise in any CFT on a sphere due to the existence of a set of conserved charges associated with conformal Killing vectors.  These charges are related to generators of an $sl(2,\mathbb R)$ algebra involving the Hamiltonian, in a manner reminiscent of the (nearly-)SU$(2)$ algebras in Sec.~\ref{sec:su(2)} and Ref.~\cite{Choi18}.  There are many open questions regarding the ubiquity and robustness of these mechanisms and the interrelations between them, if any.  At the very least, these developments suggest an exciting new direction in the study of highly excited states of quantum many-body systems and the dynamical features enabled by such states.

\acknowledgements{We acknowledge helpful discussions with D.~Abanin, A.~Chandran, W.~W.~Ho, C.-J.~Lin, and B.~Swingle. This work is supported by Microsoft and the Laboratory for Physical Sciences. T.I. acknowledges a JQI postdoctoral fellowship. S.X. acknowledges support from the U.S.~Department of Energy, Office of Science, Advanced Scientific Computing Research Quantum Algorithms Teams program as part of the QOALAS collaboration.}

\bibliography{refs_zero}

\begin{thebibliography}{58}
\bibitem [{foo()}]{foot2}%
  \BibitemOpen
  \href@noop {} {}\bibinfo {howpublished} {A possible exception to the
  criterion $\langle E|Z_\pi|m\rangle\to0$ that still allows for time-periodic
  dynamics arises in cases where $|E\rangle$ is a scar state $|m\pm n\rangle$
  with $n>1$. However, we have checked numerically that such matrix elements
  are not macroscopically large, leaving the matrix elements $\langle m\pm
  1|Z_\pi|m\rangle$ responsible for the periodic dynamics.}\BibitemShut {Stop}%
\end{thebibliography}

\end{document}